\documentclass[12pt,preprint]{aastex}

\slugcomment{submitted to the {\it Astronomical Journal}}

\shorttitle{New High Proper Motion Discoveries}
\shortauthors{Boyd et al.}

\begin{document}

\title{The Solar Neighborhood XXV: Discovery of New Proper Motion
Stars with 0$\farcs$40 yr$^{-1}$ $>$ $\mu$ $\ge$ 0$\farcs$18 yr$^{-1}$
between Declinations $-$47$\degr$ and 00$\degr$}

\author{Mark R. Boyd}
\affil{Georgia Institute of Technology, Atlanta, GA 30332}
\email{boyd@chara.gsu.edu}

\author{Jennifer G. Winters, Todd J. Henry, and Wei-Chun Jao}
\affil{Georgia State University, Atlanta, GA 30302-4106}
\email{winters@chara.gsu.edu, thenry@chara.gsu.edu, jao@chara.gsu.edu}

\author{Charlie T. Finch}
\affil{U.S. Naval Observatory, Washington DC 20392-5420}
\email{finch@usno.navy.mil}

\author{John P. Subasavage}
\affil{Cerro Tololo Inter-American Observatory, La Serena, Chile}
\email{jsubasavage@ctio.noao.edu}

\and

\author{Nigel C. Hambly}
\affil{Scottish Universities Physics Alliance, Institute for
Astronomy, University of Edinburgh, Royal Observatory, Blackford Hill,
Edinburgh EH9 3HJ, Scotland, UK}
\email{nch@roe.ac.uk}

\clearpage

\begin{abstract}

We present 2817 new southern proper motion systems with 0$\farcs$40
yr$^{-1}$ $>$ $\mu$ $\ge$ 0$\farcs$18 yr$^{-1}$ and declination
between $-$47$\degr$ and 00$\degr$.  This is a continuation of the
SuperCOSMOS-RECONS (SCR) proper motion searches of the southern sky.
We use the same photometric relations as previous searches to provide
distance estimates based on the assumption that the objects are single
main sequence stars.  We find 79 new red dwarf systems predicted to be
within 25 pc, including a few new components of previously known
systems.  Two systems --- SCR 1731-2452 at 9.5 pc and SCR 1746-3214 at
9.9 pc --- are anticipated to be within 10 pc.  We also find 23 new
white dwarf candidates with distance estimates of 15--66 pc, as well
as 360 new red subdwarf candidates.  With this search, we complete the
SCR sweep of the southern sky for stars with $\mu \ge 0$\farcs$18$
yr$^{-1}$ and $R_{59F}$ $\le$ 16.5, resulting in a total of 5042
objects in 4724 previously unreported proper motion systems.  Here we
provide selected comprehensive lists from our SCR proper motion search
to date, including 152 red dwarf systems estimated to be within 25 pc
(nine within 10 pc), 46 white dwarfs (ten within 25 pc), and 598
subdwarf candidates.  The results of this search suggest that there
are more nearby systems to be found at fainter magnitudes and lower
proper motion limits than those probed so far.

\end{abstract}


\section{Introduction}
\label{sec:intro}


The Research Consortium On Nearby Stars (RECONS)\footnote{\it
www.recons.org} has been surveying the southern sky for proper motion
objects using the database of the SuperCOSMOS Sky Survey, with new
discoveries dubbed SCR (SuperCOSMOS-RECONS).  This sixth SCR survey
paper is the second for objects with $\mu$ $\ge$ 0$\farcs$18 yr$^{-1}$
and $R_{59F}$ $\le$ 16.5, and completes our sweep of the southern sky
for objects meeting these criteria.  Results of previous efforts have
been reported in \cite{2004AJ....128..437H},
\cite{2004AJ....128.2460H}, \cite{2005AJ....129..413S,
2005AJ....130.1658S}, and \cite{2007AJ....133.2898F}.  These are
papers VIII, X, XII, XV, and XVIII in the {\it The Solar Neighborhood}
(TSN) series, respectively (hereafter TSN VIII, TSN X, etc.).  This
paper specifically complements TSN XVIII, which presented results from
our search from declinations $-$90$\degr$ to $-$47$\degr$ for objects
with proper motions, $\mu$, between 0$\farcs$40 yr$^{-1}$ and
0$\farcs$18 yr$^{-1}$.  Here we report results for objects with the
same proper motions, but for declinations $-$47$\degr$ to 00$\degr$.
We have reported a total of 1971 objects in 1907 SCR proper motion
systems in previous papers in this series, where a system is defined
to be one or more objects that appear to be gravitationally bound, as
evidenced during these searches by being near to one another on the
sky and having similar proper motions.  Here we report an additional
3073 objects in 2817 systems, which more than doubles the total SCR
count, and brings the total to 5042 objects in 4724 SCR
systems.\footnote{New proper motion systems reported in
\cite{2011AJ....141...21W} were found via customized searches
different than our typical search methodology used for this and the
other five SCR proper motion papers.  Thus, for the statistics quoted
in this paper, we only include new systems reported in
\cite{2011AJ....141...21W} if they were also (re)found via the search
methodology outlined here.}

One of the primary goals of this work is to add to the census of
stellar systems known within 25 pc, in an effort to develop the most
accurate measurements of the stellar luminosity and mass functions,
and to provide the fundamental sample of nearby stellar systems for
studies of multiplicity, activity, ages, and exoplanet searches.  To
date, our searches have focused on the southern sky, which
historically has not been searched as methodically as the northern
sky.  Our SCR efforts are, of course, the latest in a long line of
proper motion surveys for nearby stars.  Early searches include the
classics by \cite{1971lpms.book.....G,1978LowOB...8...89G} and
\cite{1979lccs.book.....L, 1980PMMin..55....1L}.  Many more have since
been done utilizing modern technology but fundamentally similar
techniques, including the searches of \cite{1994A&AS..105..179W},
\cite{1999A&AS..139...25W}, \cite{2000A&A...353..958S,
2002ApJ...565..539S}, \cite{2001Sci...292..698O},
\cite{2003A&A...397..575P, 2004A&A...421..763P},
\cite{2005AJ....130.1247L, 2008AJ....135.2177L},
\cite{2005A&A...435..363D, 2009MNRAS.397.1685D}, and
\cite{2007A&A...468..163D}.  Each of these searches has yielded new
proper motion objects, including candidates close to the Sun.  The
significance of these searches individually is discussed in detail in
previous papers in this series, so will not be addressed here.

In this paper we present photometric distance estimates for the red
dwarf, cool subdwarf, and white dwarf systems revealed during the
search.  We anticipate that 79 red dwarf systems from the present
search are within 25 pc, including two estimated to be closer than 10
pc.  While most of the systems from effort reported here are made up
of red dwarfs, we also report 23 white dwarf candidates and 360 red
subdwarf candidates selected via reduced proper motion diagrams.
These three types of intrinsically faint objects are typically
underrepresented in Galactic models.  Thus, it is important to reveal
these systems if we are to develop accurate pictures of Galactic
structure and populations.

\section{Search Criteria and Methodology}
\label{sec:criteria}

The searches use data from SuperCOSMOS scans of four Schmidt survey
photographic plates taken of each field.  The photographic plates
scanned into the SuperCOSMOS database are 6$\degr$ $\times$ 6$\degr$
with a 0.5$\degr$ overlap on each side, giving $\sim$25$\degr$ square
of unique sky coverage per field (in order to streamline computations
of astrometric and photometric data, the overlap regions were not
used). Six hundred fifty-four fields have been included in the current
search, giving a total of $\sim$16,350 square degrees covered, or
about 40\% of the entire sky.  In total, 16,748 candidate objects were
detected, which is more than twice as many as the 7410 in TSN XVIII.
TSN XV searched the same area of the sky as in this paper, albeit for
objects with $\mu$ $\ge$ 0$\farcs$40 yr$^{-1}$, and found 3879
candidate objects. The ratio of candidates from TSN XV to this paper
(0.23) is similar to that of TSN XII to TSN XVIII (0.19), which both
searched the sky from $-$47$\degr$ to the southern celestial pole with
proper motions that match TSN XV and this paper, respectively.

The current search uses techniques for object detection and extraction
of astrometric and photometric data similar to those of previous SCR
searches, which are given in detail in previous papers (see TSN XVIII
in particular).  Briefly, we utilize astrometric position and proper
motion information and photometric magnitudes in the $B_J$, $ESO-R$,
$R_{59F}$, and $I_{IVN}$ passbands.  We generally require that sources
be detected on all four plates, and that they have $R_{59F}$ $\le$
16.5 mag.  As in previous searches, an ellipticity quality flag was
checked for each of the four plates, and any object with two or more
ellipticities greater than 0.35 was excluded.  For the present search,
we added an additional check: if the mean of the three best (i.e.,
lowest) ellipticities for an object was greater than 0.25, it was
thrown out.  This cutoff was chosen because in trial samples it
removed a substantial number of false objects, but no real proper
motion objects.  Thus, it simply lowered the number of false
detections in the sample that had to be investigated.  Such
``garbage'' may result from blended images, plate defects, and objects
near bright star halos, many of which can be eliminated by these
ellipticity constraints.  Undoubtedly, a small number of true proper
motion systems were excluded using this criterion, e.g., binary
systems that were elongated, that could be picked up in future
searches with relaxed criteria.  In addition to the ellipticity
constraints, the further sifting process discussed in TSN XII and
applied to subsequent searches was also used here --- if the two {\it
R} magnitudes differed by more than 1.0 magnitude the object was
considered to be a mismatch or a variable giant with an erroneous
proper motion, and was discarded from further consideration.

Once a list of reliable candidates was extracted, the objects were
then checked using SIMBAD and other proper motion surveys [e.g., NLTT
  --- \cite{1980nltt.book.....L}, LEHPM ---
  \cite{2003A&A...397..575P}, SIPS --- \cite{2005A&A...435..363D}] in
VizieR to identify previously known objects.  In both SIMBAD and
VizieR, a 90\arcsec~radius was used to match objects in accordance
with the findings of \cite{2002ApJS..141..187B}, who found that
coordinates of stars in the Luyten half-second (LHS) catalog were
usually accurate to within $\sim$90$\arcsec$ (see their Figure 2).  All
new and known candidate proper motion objects were inspected using
Aladin by blinking the $B_J$, $R_{59F}$, and $I_{IVN}$ SuperCOSMOS
plate images to confirm that the object was indeed a proper motion
object.  For real objects, 2MASS positions, epochs, and $JHK_s$
photometry were extracted.  The blinking was done using a
5\arcmin~radius field and SIMBAD and VizieR overlays to ensure new
discoveries were not incorrectly labeled as known and vice versa, and
to ensure the correct 2MASS data were collected.  The blinking process
led to the discovery of many common proper motion systems, discussed
in $\S$\ref{sec:cpm}.


%

\section{Comparison to Previous Searches}
\label{sec:previoius}
As in previous papers, we examine the discovery statistics of our SCR
search, which have been updated to include systems from this paper.
Results are summarized in Table~\ref{discostats}.  In order to be
consistent with TSN XVIII, we reintroduce the terminology used there
to describe the various samples. We divide the systems into three
categories: MOTION, SLOWMO, and MINIMO, which contain systems with
$\mu$ $\ge$ 1$\farcs$00 yr$^{-1}$, 1$\farcs$00 yr$^{-1}$ $>$ $\mu$
$\ge$ 0$\farcs$50 yr$^{-1}$, and 0$\farcs$50 yr$^{-1}$ $>$ $\mu$ $\ge$
0$\farcs$18 yr$^{-1}$, respectively, with the two lower cutoffs chosen
to match those of the Luyten Half Second (LHS) and Luyten Two Tenths
(LTT) efforts.  Here we update the hit rates --- defined as the number
of real objects, including new, known, and duplicates of new and known
objects, divided by the total starting sample size including all
candidate objects --- for the MOTION, SLOWMO, and MINIMO samples from
those given in TSN XVIII, which included SCR searches only between
$-$90$\degr$ to $-$47$\degr$.  For the entire southern sky SCR
searches, we find hit rates of 8.9\% and 80.2\% for the MOTION and
SLOWMO samples, respectively.  The previous hit rate for MINIMO
systems, the focus of this paper, between $-$90$\degr$ to $-$47$\degr$
was 78.1\%.  In this paper our hit rate is 81.4\%, which is slightly
higher because of the additional ellipticity constraint that
eliminated $\sim$500 garbage entries.\footnote{The hit rate for all
MINIMO objects listed in Table~\ref{discostats} is 81.5\%, which is
slightly higher than either of the rates given for the searches in TSN
XVIII and that reported here, which include objects with 0$\farcs$40
yr$^{-1}$ $>$ $\mu$ $\ge$ 0$\farcs$18.  Objects with proper motions of
0$\farcs$50 yr$^{-1}$ $>$ $\mu$ $\ge$ 0$\farcs$40 have very high,
i.e., reliable, hit rates, thereby increasing the overall hit rate for
MINIMO systems.}  Table~\ref{discostats} lists the distribution of
real and garbage objects divided into their appropriate proper motion
bins.

There are 13,363 objects in the NLTT catalog that meet the sky
location, magnitude, and proper motion parameters of the search
reported here.  Of these, 9474 were recovered, corresponding to a 71\%
recovery rate.  The 29\% not recovered can be attributed primarily to
the factors mentioned in TSN XVIII --- differences in proper motion
and magnitude measurements can lead to us dropping objects that were
kept in the NLTT, i.e., the SCR measurements are different enough to
push objects beyond the designated search limits.  In addition, our
search has trouble picking up bright sources. The brightest NLTT
object recovered has $R_{59F}$ $\sim$ 5, while the brightest NLTT
object in this part of the sky has Luyten's $r$ $\sim$ 2.

\section{Data}
\label{sec:data}
  
In Table~\ref{minimofive} we list the first five of the 2817 total
systems discovered during the present search.  The table of
discoveries is presented in full in the electronic version of the {\it
Astronomical Journal}.  For this total, we only count systems
comprised entirely of new discoveries, i.e., an SCR companion to a
known object is {\it not} included in this number. There are 3073
total SCR objects from the present search, which exceeds the number of
systems because (a) some systems have more than one SCR object and (b)
systems including both a known and SCR object are not included in the
number of systems.  We provide SCR names, coordinates, relative proper
motions and position angles of the proper motions, plate magnitudes
from SuperCOSMOS, photometry from 2MASS, the $R_{59F} - J$ color, a
distance estimate, and notes, as we have done for systems found in
previous SCR searches. The proper motions and position angles have
errors of $\sim$ 0$\farcs$010 yr$^{-1}$ and $\sim$ 2.7$\degr$,
respectively. All coordinates have been computed for epoch J2000.0
using the 2MASS coordinates and the SuperCOSMOS proper motions and
position angles.  Tables~\ref{allscr-25}, \ref{subdwarffive}, and
\ref{allscr-wd} provide summary lists of our discoveries of red dwarfs
within 25 pc, cool subdwarfs, and white dwarfs, respectively, from the
searches to date using the methodology outlined in the six SCR proper
motion papers.


In TSN XVIII, the proper motion and position angle data from the SCR
searches were shown to be consistent with those of {\it Hipparcos} and
the NLTT. {\it Hipparcos} observed stars brighter than {\it V} $\sim
12$, which tend to have the poorest proper motions in the SCR survey
because of image saturation on the photographic plates.  Even so, the
proper motions and position angles had average deviations of
0$\farcs$020 yr$^{-1}$ and 3.9$\degr$, respectively.  The agreement
between SCR and NLTT values are rather worse, at 0$\farcs$025
yr$^{-1}$ and 6.8$\degr$.  \cite{2010AJ....140..844F} compared UCAC3
[\cite{2010AJ....139.2184Z}] and SuperCOSMOS proper motion data for
137 objects in both catalogs. The average differences found were less
than 0$\farcs$020 yr$^{-1}$. We also compare the SCR sample to the
PPMX [\cite{2008A&A...488..401R}] and PPMXL
[\cite{2010AJ....139.2440R}] catalogs. We have searched a single hour
of RA between RA $=$ 12 and 13 to derive representative samples for
comparison, with the search radii for object matching set at 30
arcseconds. In the PPMX catalog, we recovered 52 of 158 SCR objects,
or 33\%, with the proper motions differing by an average of
0$\farcs$021 yr$^{-1}$ and 0$\farcs$027 yr$^{-1}$ for RA and Dec,
respectively. In the PPMXL catalog, we recovered 115 of 158 SCR
objects, or 73\%, with the proper motions differing by an average of
0$\farcs$024 yr$^{-1}$ and 0$\farcs$027 yr$^{-1}$ for RA and Dec,
respectively. These values are the mean differences between proper
motions values in RA and Dec for the catalogs.  Systematic offsets
between the SCR and PPMX results in RA and Dec are $-$0$\farcs$022
yr$^{-1}$ and 0$\farcs$013 yr$^{-1}$, whereas offsets between SCR and
PPMXL results are $-$0$\farcs$026 yr$^{-1}$ and 0$\farcs$011
yr$^{-1}$. These indicate a small systematic shift between the
catalogs.

\cite{2004AJ....128..437H} describe the 11 relations generated from
the six photometry values, $B_JR_{59F}I_{IVN}JHK_s$ (hereafter
$BRIJHK$), associated with each object that can be used to estimate
distances photometrically.  The relations were generated using stars
with accurate (errors less than 10 mas) trigonometric parallaxes, and
the estimates assume that each object is a single main sequence star
with colors corresponding to a K or M dwarf.  Stars for which fewer
than six relations produced distance estimates are noted with
distances in brackets; these objects typically have colors too blue
for the relations.  For the stars with accurate trigonometric
distances used to generate the relations, the mean of the absolute
differences between their true distances and their estimated
photometric distances is 26\%.  Thus, errors on the distance estimates
listed for red stars are a minimum of 26\%.  An additional error for
an object's distance arises from the standard deviation of results
from the (up to) 11 different distances from the relations.  Away from
the Galactic plane, this standard deviation error is typically 15\%,
which results in total errors for the distances of $\sim$30\%.  Near
the Galactic plane, where crowding is an issue and one or more images
on the plates may be corrupted by background sources, the photometry
is less accurate and consistent between the $BRI$ plate magnitudes, so
errors may climb to 50\% or more in extreme cases.

Distance estimates will be erroneous for certain classes of stars,
notably white dwarfs and subdwarfs, both of which are underluminous
compared to main sequence stars of the same color.  Thus, the
overestimated distances for these candidates are listed in brackets in
Table~\ref{minimofive}.  Where possible, white dwarf candidates have
more accurate distances listed in the notes of Table~\ref{allscr-wd}.
SuperCOSMOS data were gathered manually for companions noticed by eye
during the blinking process that appeared to have common proper motion
with a target being checked.  These objects were typically not picked
up during the initial search because they are fainter than the $R$ $=$
16.5 cutoff.  Some lack SuperCOSMOS data, and therefore distance
estimates, because they are blended on the plates or are too faint for
reliable SuperCOSMOS magnitudes.

We anticipate a roughtly 1\% contamination rate of false positive
proper motion objects in the sample of 2817 systems.  Some false
positives are particularly bright stars on the B or I plates where the
blinker must use diffraction spikes to judge the location of the image
center because the image itself is large and/or asymmetric.  In these
cases, we erred on the side of inclusion, so the list likely contains
a a few bright objects that do not really exhibit proper motion.  A
second type of false positives are objects near plate edges.  Plate
edges are effectively "cut" at designated locations to provide full
sky coverage.  In some cases, individual star images are split
between, for example, a B plate and an R plate, and those images are
offset, thereby causing a false proper motion.  These detections were
kept as candidates because they meet the SCR search criteria, but may
not to be proper motion objects.


\section{Analysis}
\label{sec:analysis}

\subsection{Color-Magnitude Diagram}
\label{sec:cmd}


Figures~\ref{cmdnew} and~\ref{cmdknown} show color-magnitude diagrams
for SCR objects and previously known systems recovered during the
present search.  As in previous SCR efforts, the systems discovered
here are generally fainter and redder than previously known systems,
with a concentration of points around $R$ $\sim$ 15 and $(R-J) \sim$
3, representing red dwarfs of spectral types $\sim$M2.0V.  The 21 SCR
objects with $(R-J) \ge$ 4.5 are estimated to have spectral types of
M5.0V to M8.0V.  As in TSN XVIII, many of the systems discovered are
brighter and bluer than in earlier SCR search papers.  There are 55
objects with $R$ brighter than 10, with SCR1843-0146 at $R =$ 8.54 mag
being the brightest.  TSN XVIII reported only nine objects brighter
than $R =$ 10.  In addition to sample size, the difference in
detection rates may be due to the omission during the TSN XVIII search
of several plates near the Galactic plane, where blue proper motion
objects may be found superimposed on the swath of background stars and
dust of the plane.  Because of their blue colors, none of these
objects have reliable distance estimates presented here, as the suite
of distance estimate equations is only applicable to red stars.
Nonetheless, the stars are bright and have confirmed proper motions so
are certainly worthy of follow-up work.  Points below the search
cutoff of $R=$ 16.5 represent common proper motion companions noticed
by eye during the blinking process.  In addition, white dwarf
candidates represented by triangles are immediately visible, clustered
around $R \sim$ 16 and $(R-J) \sim$ 0.  The subdwarf population, while
less well-defined, is also noticeable as a group of points stretching
from $R \sim$ 12 and $(R-J) \sim$ 0 to $R \sim$ 16 and $(R-J) \sim$ 2.
Further assessment of white dwarf and cool subdwarf candidates can be
accomplished using a reduced proper motion diagram.

\subsection{Reduced Proper Motion Diagram}
\label{sec:rpmd}

Figure~\ref{rpmd} shows the reduced proper motion (RPM) diagram for
SCR objects found during the present survey.  The RPM diagram is a
powerful tool for estimating the luminosity class of a star. It is
similar in nature to the H-R diagram except that the proper motion is
used instead of a distance, relying on the inverse statistical
relationship between proper motion, $\mu$, and distance. While
obviously not foolproof --- for example, subdwarfs may masquerade as
main sequence stars, and vice versa --- the diagram allows for the
rough classification of systems.  The equation used here to determine
the pseudo-absolute magnitude $H_{R_{59F}}$ plotted on the vertical
axis of Figure~\ref{rpmd} is

\begin{displaymath}
H_{R_{59F}} = R_{59F} + 5 + 5\log\mu.
\end{displaymath}
  
The dashed line in Figure~\ref{rpmd} is the same as for similar RPM
plots in TSN XII, XV, and XVIII, and is used to separate white dwarfs
from subdwarfs.  The solid lines, while not plotted in previous
papers, are used to denote the cool subdwarf area on the plot. We find
23 new white dwarf candidates from this effort.  As in previous
searches, we also use the RPM diagram to identify cool subdwarf
candidates.  From this search, there are 360 subdwarf candidates
defined as having $R-J > $ 1.0 and $H_{R_{59F}}$ up to 4.0 mag
brighter than the white dwarf-subdwarf line.  Although this is a
somewhat arbitrary definition, it has proven reliable for delineating
subdwarfs from both main sequence stars and white dwarfs.  Samples
selected from the red dwarf, cool subdwarf, and white dwarf regions of
the RPM diagram are discussed in the next three sections.

\subsection{SCR Red Dwarfs Within 25 Parsecs}
\label{sec:nearby}


Table~\ref{allscr-25} lists the 152 red dwarf systems estimated to be
within 25 pc found during the SCR proper motion surveys.  We provide
coordinates, proper motions, plate and 2MASS magnitudes, a photometric
distance estimate using the suite of relations presented in TSN VIII,
the paper in which the object was first published, and notes.  Nine of
the systems are estimated to be in the RECONS 10 pc sample.  We
reported trigonometric parallaxes for three of the systems in
\cite{2006AJ....132.2360H}, including SCR 0630-7643AB at 8.76 pc, SCR
1138-7721 at 8.18 pc, and SCR 1845-6357AB at 3.85 pc.  As of January
1, 2011, the latter system ranked as the 23rd nearest system to the
Sun\footnote{\it www.recons.org/TOP100.posted.htm}.

The current search adds 79 systems to the 25 pc sample, more than
doubling the number of systems within this volume from previous SCR
search efforts, including two systems likely to be within 10 pc ---
SCR 1731-2452 at 9.5 pc and SCR 1746-3214 at 9.9 pc.  Four of the nine
objects estimated to be within 10 pc have proper motions greater than
$0\farcs50$ $yr^{-1}$, the cutoff of the Luyten Half Second (LHS)
catalog, while the other five are moving more slowly.  For objects
between 10 and 25 pc, 118 of 144 have proper motions less than
$0\farcs50$ $yr^{-1}$.  The fact that so many nearby objects have
relatively low proper motions suggests that there may yet be more
nearby systems at even lower proper motions.

\subsection{SCR Cool Subdwarfs}
\label{sec:subdwarf}

During the SCR surveys we have also identified potential cool
subdwarfs, including 598 total candidates, of which 360 are from this
paper.  We provide data similar to that given for the SCR red dwarfs
for the first five entries in Table~\ref{subdwarffive}; the complete
list is given in the electronic version of the {\it Astronomical
Journal}.

Subdwarfs are less luminous than their main-sequence counterparts and
so have distance estimates that are larger than their true distances.
For this reason, their distances have been listed in brackets in
Tables~\ref{minimofive} and~\ref{subdwarffive}.  The methodology used
to identify the subdwarfs, detailed in $\S$\ref{sec:rpmd}, leads to
some contamination of the sample by white dwarfs and main sequence
objects, so spectroscopic confirmation is desired.  We have continuing
programs to confirm cool subdwarfs spectroscopically
\citep{2008AJ....136..840J} and to measure their distances via
trigonometric parallax, as described in \citet{2005AJ....129.1954J}
and Jao et al.~(2011), the latter reporting the first two SCR subdwarf
parallaxes, for SCR 1107-4135 (67.61 pc) and SCR 1916-3638 (67.66 pc).


\subsection{SCR White Dwarfs}
\label{sec:wd}

During the SCR surveys we have found 46 white dwarf candidates, which
are listed in Table~\ref{allscr-wd}.  All of the white dwarf
candidates were selected based on the criteria described in
$\S$\ref{sec:rpmd}.  The current search has provided 23 white dwarf
candidates, matching the total of our previous SCR searches.  In
Table~\ref{allscr-wd}, we provide the same data as for the 25 pc red
dwarf and subdwarf tables, except the listed distance estimates are
from the single color linear relation of \cite{2001Sci...292..698O}.
Ten white dwarf candidates are estimated to be within 25 pc using this
relation, although none have been added to the 10 pc
sample.\footnote{We note that the potentially nearest white dwarf
candidate, SCR 1800-5112B at 10.2 pc from TSN XVIII, has colors that
are inconsistent with it actually being a white dwarf.  Nonetheless,
we include this object in Table~\ref{allscr-wd} because it falls below
the white dwarf cutoff line drawn in the RPM diagram.}

Fifteen of the 46 candidates discovered have been spectroscopically
confirmed to be white dwarfs by
\cite{2007AJ....134..252S,2008AJ....136..899S}, and similar
spectroscopic confirmation is underway for the rest of the SCR white
dwarf sample.  Of particular interest are three relatively hot WDs
($T_{\rm eff}$ $>$ 10,000 K) estimated to be within 25 pc --- SCR
1920-3611 at 14.7 pc, SCR 1107-3420A at 16.0 pc, and SCR 0711-2518 at
20.3 pc (see $\S$\ref{sec:individual}).  The distance relation of
\cite{2001Sci...292..698O} becomes unreliable at hotter effective
temperatures, in effect, because hot white dwarfs are underrepresented
in the local neighborhood sample of white dwarfs that was used to
generate the relation.  Thus, we adopt the distance estimates
determined by \cite{2007AJ....134..252S,2008AJ....136..899S} using
$VRIJHK_s$ and fits to atmospheric models.  These distance estimates,
as well as those for all of the spectroscopic confirmations, are
listed in the notes of Table~\ref{allscr-wd}.

In \cite{2009AJ....137.4547S}, we reported trigonometric parallaxes
for four of these white dwarfs, SCR 0753-2524 (17.69 pc), SCR
0821-6703 (10.65 pc), SCR 2012-5956 (16.55 pc), and SCR 2016-7945
(24.96 pc).  Using the best distance estimates available, we find
seven WDs predicted to be within 25 pc, Of these, three have proper
motions less than 0$\farcs$50 yr$^{-1}$, while the 39 beyond 25 pc all
have proper motions less than 0$\farcs$50 yr$^{-1}$.

\subsection{Common Proper Motion Systems}
\label{sec:cpm}

This search yielded 250 potential common proper motion (CPM) systems,
listed in Table~\ref{cpm-table}.  These systems were found in two
ways.  First, if two sources appeared to be CPM when blinking frames,
they were noted for further investigation.  Second, we used search
criteria that linked pairs of objects with separations $\leq$
1200\arcsec, $\Delta$$\mu$ $\leq$ 0$\farcs$025 yr$^{-1}$, and
$\Delta$$\theta$ $\leq$ 15$\degr$.  In total, 121 systems have all
components as new discoveries, and an additional 129 systems have at
least one new SCR component.  The list includes 239 doubles, ten
triples, and one possible quintuple system.  Table~\ref{cpm-table}
lists identifiers for system members, the proper motions and position
angles for each component, separations and position angles between the
components, distance estimates from the $BRIJHK$ photometry where
available, and notes.  The separations and position angles were
determined using 2MASS positions and used spherical trigonometric
relations.  As in TSN XVIII, because of errors in the plate relations,
distance estimates that agree to within a factor of two are considered
to indicate a candidate system.  Table~\ref{cpm-table} is broken into
two classes of CPM candidate systems depending on the reliability of
physical association.  Those at the top we consider probable because
all components of the systems have complete sets of $\mu$ and $\theta$
values that match within 0$\farcs$025 yr$^{-1}$, and 15$\degr$.  The
second class of systems have either (a) mismatched $\mu$ and $\theta$
values extracted from SuperCOSMOS, or (b) no available values.  In the
former case, the pairs often appeared to be better matched when
blinking the plate images, and the available data are suspected to be
erroneous.  Companions that were not retrieved during the initial
search tended to be either fainter than the $R_{59F}$ $=$ 16.5 cutoff
or moving slightly beyond the limits of the proper motion range
searched.

Figures~\ref{mumu} and~\ref{papa} compare the proper motion sizes and
position angles for components of multiple systems, respectively.  As
is well known, the position angle of an object's proper motion is
often better determined than the size of its proper motion,
particularly for low proper motions.  As such, the position angles are
more reliable in helping determine whether or not two moving objects
comprise a system.  Systems for which at least one component had its
proper motion and position angle data gathered manually from the
SuperCOSMOS Sky Survey database are marked with open circles. These
data were retrieved from the SuperCOSMOS Sky Survey website one-by-one
and tend to be less reliable than those from the initial search
because the initial search utilized a specialized high proper motion
version of the SuperCOSMOS data. The SSS website from which some data
were retrieved used nearest-neighbor pairing within a restricted
radius, while the specialized database used all possible pairings
regardless of spurious pairings (which were removed at a later
stage). Systems marked with solid points are those with both
components retrieved during the initial automated search, and tend to
have better agreement between these values, particularly for the
proper motions. There is more spread in the distribution of points in
Figures~\ref{mumu} and~\ref{papa} here that compare components in
multiple systems than is seen in comparable Figures 6 and 7 in
\cite{2010AJ....140..844F}. The reason is that in order to be
circumspect in our search for possible multiple systems, we have
relaxed the by-eye criteria in this paper.


\subsection{Sky Distribution of SCR Systems}
\label{sec:allscr-distro}

Figure~\ref{distro} shows the sky distribution of SCR systems, broken
into discoveries from the first five proper motion papers and this
paper.  Of note for this portion of the survey is the dense bar from
RA $=$ 4h to 22h and declinations $-$33$\degr$ to $-$47$\degr$, which
fills in the sparse area in Figure~\ref{nlttdistro}, the southern sky
distribution of NLTT systems.  A dense patch from RA $=$ 6h to 9h
traces the Galactic plane (represented by a solid line), which has
previously been poorly searched in the southern hemisphere.  The area
from RA $=$ 17h to 19h centered near declination $-$35$\degr$
corresponds to the Galactic center/bulge region, which is an
extraordinarily crowded region on the plates and in the SuperCOSMOS
database.  Our detection rate is much lower there than in the rest of
the Galactic plane, where the crowding is not so extreme.  However,
some systems have been discovered in this region because they have a
backdrop of gas and dust that obscures many of the background stars
that would otherwise make unique source identification difficult.

\subsection{Comments on Individual Systems}
\label{sec:individual}

Here we highlight a few of the 2817 systems reported from this portion
of the SCR search.  There are many systems worthy of further
investigation, but those selected here represent the most extreme
cases in proximity, color, brightness, or complexity.  Trigonometric
parallax observations are underway for many of these targets.

{\bf SCR 0225-1829AB} is a binary system in which the primary is the
bluest object found in the present search, having $R-J = -$1.42. It is
also one of the brightest, with $R =$ 8.63. As such, a distance
estimate is unavailable. The secondary, however, is a red dwarf
estimated to be at 28.7 pc.

{\bf SCR 0711-2518} is a hot white dwarf initially estimated to be at
20.3 pc using the single-color photometric distance estimate.
However, this is inaccurate because it is a very hot white dwarf.
\cite{2008AJ....136..899S} presents spectroscopic confirmation and an
updated distance estimate of 30.5 pc using CCD photometry and
atmospheric models, thus likely placing it beyond the 25 pc horizon.

{\bf SCR 1107-3420AB} is in a binary system with a red dwarf estimated
to be at 19.2 pc. \cite{2007AJ....134..252S} reported spectroscopic
confirmation of the white dwarf.  The initial photometric distance
estimate of 16.0 pc is inaccurate because the white dwarf is hot.
Thus, we adopt the distance estimate of 28.2 pc presented by
\cite{2007AJ....134..252S} using CCD photometry and atmospheric
models. The red dwarf's distance estimate leaves open the possibility
that this system lies within 25 pc.

{\bf SCR 1337-1046DE} is a blended pair and part of a possible
quintuple system that includes NLTT 34652 as the primary and the
double HD 118512BC.  We are obtaining CCD photometry to help confirm
or refute the nature of this potentially complex system.

{\bf SCR 1731-2452} ($R =$ 13.39, $R-J =$ 4.12) is the nearest star
from this portion of the SCR search, a red dwarf with an estimated
distance of 9.5 pc.

{\bf SCR 1746-3214} ($R =$ 15.89, $R-J =$ 5.56) is the second nearest
star from this portion of the SCR search, a red dwarf with an
estimated distance of 9.9 pc.  It is also one of the reddest objects
found.

{\bf SCR 1843-0146} is the brightest object found in this search at $R
=$ 8.54. With $R-J = -$0.14, it is too blue for us to estimate a
reliable distance using the suite of 11 plate relations.

{\bf SCR 1920-3611} is a hot white dwarf initially estimated to be
within 25 pc (14.7 pc). This estimate is inaccurate because the white
dwarf is too hot for the relation.  \cite{2008AJ....136..899S}
presents spectroscopic confirmation and an updated distance estimate
of 41.7 pc using CCD photometry and atmospheric models.

{\bf SCR 2354-0352AB} is a CPM system that has both exceptionally good
agreement on distance estimates (A $=$ 60.6 pc, B $=$ 60.0 pc) and a
very wide separation, 695$\farcs$4.  This gives a projected separation
of $\sim$ 42000 AU for the pair, making it an extraordinary candidate
for a very wide multiple system.

\section{Discussion}
\label{sec:discussion}

This paper completes our SCR sweep of the entire southern sky for
objects with ${\it R_{59F}}$ $\le$ 16.5 and $\mu$ $\ge$ 0$\farcs$18
yr$^{-1}$ using the methodology outlined.  In total, the SCR search
has revealed 4724 new proper motion systems, 2817 of which are from
the effort described in this paper.  Among these discoveries, we have
found 152 red dwarf systems within 25 pc, including nine within 10 pc.
Seventy-nine of the 25 pc systems, including two of the 10 pc systems
were revealed in the current search.  In addition to these red dwarf
systems, 46 white dwarf candidates (ten estimated to be within 25 pc)
and 598 cool subdwarf candidate systems have been found, with 23 white
dwarf candidates and 360 candidate subdwarfs from the current search.
Overall, the total of 5042 objects added via the SCR searches
constitutes an increase of $\sim$20\% over the number of entries in
the NLTT south of DEC = 0$\degr$.

Table~\ref{diststats} provides the discovery statistics for the SCR
sample to date, broken into bins by proper motion and distance.  The
first number in each column represents the number of systems from
previous papers, and the second is the number of new systems from this
paper.  The breakdown confirms the trend described in TSN XVIII that
slower proper motion bins have comparable numbers of 10 pc objects
(although the small numbers of objects in each bin preclude a robust
analysis) and many more 25 pc objects.  This suggests that there exist
more systems very near the Sun at even lower proper motions.  The
nearest systems also tend to be among the faintest and reddest of the
red dwarfs.  Many of the 25 pc objects, and four of the nine 10 pc
objects, have $R$ within 1.0 mag of our cutoff of 16.5 mag.  This
suggests that there are many nearby objects at fainter magnitudes than
those we have probed so far.  We are currently conducting searches for
both slower proper motion and fainter objects to reveal additional
nearby systems.  Additional searches using both the SuperCOSMOS and
other databases will undoubtedly reveal new proper motion systems, as
has already been done using UCAC3 in a complementary search of TSN
XVIII by \cite{2010AJ....140..844F}.

The SCR survey sample has provided a rich sample of CPM systems, many
of which have very wide separations.  The sky separations range from
less than five to nearly 700 arcseconds (e.g., Table~\ref{cpm-table}).
Even at relatively close separations, the estimated distances to the
systems give projected sky separations of hundreds of AU.  The most
widely separated systems have extremely large spatial separations,
often greater than 10,000 AU.  Because most of the systems found by
the SCR searches contain red dwarfs of low mass, their binding
energies are quite low.  We are currently gathering CCD photometry for
a sample of the widest pairs to provide better distance estimates and
to determine whether or not these systems are, indeed, gravitationally
bound.  If so, these wide pairs can be used to explore the long-term
survival of such systems, and provide clues about the overall mass
content of the Galaxy.  In addition, because we have found several
hundred multiple systems of low mass with reasonably accurate distance
estimates, we can begin to map out the distributions of separations
and mass ratios that result from the star formation process.

Finally, we are measuring accurate trigonometric parallaxes for
$\sim$60 of the SCR systems as part of our Cerro Tololo Inter-American
Observatory Parallax Investigation (CTIOPI), an astrometry program
carried out at the CTIO 0.9m [\cite{{2005AJ....129.1954J}},
\cite{2006AJ....132.2360H}, \cite{2009AJ....137.4547S}, and
\cite{2010AJ....140..897R}].  We are focusing primarily on the nearest
and highest proper motion systems.  While we cannot observe the
complete samples of nearby red dwarfs, subdwarfs, and white dwarfs
found during the SCR search, we hope that by identifying these
potentially nearby stars now, we will be poised to take advantage of
the future large scale parallax efforts such as Gaia and LSST, and
thereby help to paint a more accurate portrait of the solar
neighborhood.


\section{Acknowledgments}

The RECONS effort is supported by the National Science Foundation
through grant AST 09-08402.  This effort has also been supported by
NASA grant NNX08BA95G.  We thank GSU Honors Students Gregory Brooks,
Daryl Giuliano, Skyler Green, and Scott Stinson, each of whom assisted
in the blinking process.  Funding for the SuperCOSMOS Sky Survey was
provided by the UK Particle Physics and Astronomy Research Council.
N.C.H.~would like to thank colleagues in the Wide Field Astronomy Unit
at Edinburgh for their work in making the SSS possible; particular
thanks go to Mike Read, Sue Tritton, and Harvey MacGillivray.  This
research has made use of results from the SAO/NASA Astrophysics Data
System Bibliographic Services, the SIMBAD and VizieR databases
operated at CDS, Strasbourg, France, and the Two Micron All Sky
Survey, which is a joint project of the University of Massachusetts
and the Infrared Processing and Analysis Center, funded by NASA and
NSF.

\clearpage


\clearpage


\hoffset-00pt{}
\begin{deluxetable}{lrrrr}
\tablecaption{Discovery Statistics for Entire SCR Sample to Date\tablenotemark{a}
\label{discostats}}
\tablewidth{0pt}

\tablehead{\vspace{-8pt} \\
\colhead{Category}&
\colhead{MOTION\tablenotemark{b}}&
\colhead{SLOWMO\tablenotemark{c}}&
\colhead{MINIMO\tablenotemark{d}}&
\colhead{Total}}

\startdata
New Discoveries      &    9 &  141 &  4574 &  4724 \\ 
Known                &  171 & 1159 & 17244 & 18574 \\ 
Duplicates           &   15 &   91 &  1640 &  1746 \\ 
Garbage              & 1989 &  344 &  5335 &  7668 \\ 
  Total hits         & 2184 & 1735 & 28793 & 32712 \\ 
\enddata

\tablenotetext{a}{A few objects were also reported in searches done concurrently by
\cite{2005A&A...435..363D,2007A&A...468..163D,2005AJ....130.1247L,2008AJ....135.2177L}.}
\tablenotetext{b}{MOTION sample includes $\mu$ $\ge$ 1$\farcs$00 yr$^{-1}$ }
\tablenotetext{c}{SLOWMO sample includes 1$\farcs$00 yr$^{-1}$ $>$ $\mu$ $\ge$ 0$\farcs$50 yr$^{-1}$}
\tablenotetext{d}{MINIMO sample includes 0$\farcs$50 yr$^{-1}$ $>$ $\mu$ $\ge$ 0$\farcs$18 yr$^{-1}$}
\end{deluxetable}

\clearpage


\begin{deluxetable}{lcccccrrrrrrrl}
\rotate \tabletypesize{\scriptsize} \tablecaption{New SCR Objects with
  0$\farcs$40 yr$^{-1}$ $>$ $\mu$ $\ge$ 0$\farcs$18 yr$^{-1}$ between
  Declinations $-$47$\degr$ and 00$\degr$
\label{minimofive}}
\tablewidth{0pt}

\tablehead{\vspace{-8pt} \\
\colhead{Name}&
\colhead{RA}&
\colhead{DEC}&
\colhead{$\mu$}&
\colhead{$\theta$}&
\colhead{$B_J$}&
\colhead{$R_{59F}$}&
\colhead{$I_{IVN}$}&
\colhead{$J$}&
\colhead{$H$}&
\colhead{$K_s$}&
\colhead{$R_{59F}$ $-$ $J$}&
\colhead{Est Dist}&
\colhead{Notes}\\

\colhead{}&
\colhead{(J2000)}&
\colhead{(J2000)}&
\colhead{($\arcsec$)}&
\colhead{($\degr$)}&
\colhead{}&
\colhead{}&
\colhead{}&
\colhead{}&
\colhead{}&
\colhead{}&
\colhead{}&
\colhead{(pc)}&
\colhead{}}

\startdata
\vspace{0pt} \\


SCR 0000-4117B & 00 00 19.18 & -41 17 46.2 & 0.245 &  124.1  &\nodata & 20.26  &  17.45 &  15.00 &  14.47 &  14.48 & 5.26 &  114.9   & \tablenotemark{a} \\
SCR 0001-2641  & 00 01 37.84 & -26 41 51.8 & 0.184 &  142.1  &  14.44 & 13.39  &  12.95 &  12.53 &  12.12 &  12.00 & 0.86 & [147.5]  & \tablenotemark{b} \\
SCR 0002-0622  & 00 02 31.46 & -06 22 49.5 & 0.187 &  117.4  &  12.03 & 11.10  &  10.40 &  10.16 &  9.66  &  9.57  & 0.94 &  [55.1]  & \tablenotemark{b} \\
SCR 0002-4644A & 00 02 35.66 & -46 44 51.9 & 0.197 &  122.5  &  17.20 & 15.39  &  14.70 &  14.23 &  13.60 &  13.45 & 1.16 & [328.3]  & \tablenotemark{ac}\\
SCR 0002-4644B & 00 02 36.20 & -46 44 57.8 & 0.240 &  112.3  &  21.82 & 19.48  &  18.33 &  16.95 &  16.17 &  15.92 & 2.53 & [625.3]  & \tablenotemark{ac}\\
\enddata
                             
\tablenotetext{a}{Common proper motion system; see Table~\ref{cpm-table}.}
\tablenotetext{b}{Fewer than six relations for distance estimate, therefore unreliable [in brackets].}
\tablenotetext{c}{Subdwarf candidate, see Table~\ref{subdwarffive}; unreliable distance estimate [in brackets].}
\tablenotetext{d}{White dwarf candidate, see Table~\ref{allscr-wd}; unreliable distance estimate [in brackets].}
\tablenotetext{e}{Proper motion or position angle suspect.}
\tablenotetext{f}{No 2MASS data available, so no distance estimate.}
\tablenotetext{g}{Coordinates not J2000.0 due to lack of proper motion or 2MASS data. SuperCOSMOS coordinates used instead.}
\tablenotetext{h}{Red dwarf candidate within 25 pc, see Table~\ref{allscr-25}.}

\end{deluxetable}

\clearpage


\begin{deluxetable}{lcccccccccccll}
\rotate \tabletypesize{\scriptsize} \tablecaption{SCR Red Dwarf
  Candidates Estimated to be within 25 Parsecs
\label{allscr-25}}
\tablewidth{0pt}

\tablehead{\vspace{-8pt} \\
\colhead{Name}&
\colhead{RA}&
\colhead{DEC}&
\colhead{$\mu$}&
\colhead{$\theta$}&
\colhead{$B_J$}&
\colhead{$R_{59F}$}&
\colhead{$I_{IVN}$}&
\colhead{$J$}&
\colhead{$H$}&
\colhead{$K_s$}&
\colhead{Est Dist}&
\colhead{TSN}&
\colhead{Notes}\\

\colhead{}&
\colhead{(J2000)}&
\colhead{(J2000)}&
\colhead{($\arcsec$)}&
\colhead{($\degr$)}&
\colhead{}&
\colhead{}&
\colhead{}&
\colhead{}&
\colhead{}&
\colhead{}&
\colhead{(pc)}&
\colhead{}&
\colhead{}}

\startdata
\vspace{0pt} \\

SCR 0017-3219      & 00 17 15.73 & -32 19 54.0 & 0.210 & 094.3 & 16.81 & 14.69 & 12.48 & 10.64 & 10.08 &  9.73 & 20.6 & XXV   & 21.3 pc\tablenotemark{a}\\
SCR 0027-0806      & 00 27 45.36 & -08 06 04.7 & 0.184 & 122.3 & 18.65 & 16.26 & 13.49 & 11.57 & 10.97 &  10.61& 22.9 & XXV   & 18.6 pc\tablenotemark{a}\\
SCR 0111-4908      & 01 11 47.51 & -49 08 09.0 & 0.542 & 213.1 & 18.93 & 16.50 & 13.01 & 11.54 & 11.00 & 10.61 & 23.6 & XII   & 17.6 pc\tablenotemark{a}\\
SCR 0135-6127      & 01 35 53.66 & -61 27 11.1 & 0.255 & 256.8 & 15.62 & 13.67 & 11.81 & 10.06 &  9.53 &  9.24 & 20.8 & XVIII & 24.8 pc\tablenotemark{a}\\      
SCR 0137-4148      & 01 37 23.49 & -41 48 56.2 & 0.238 & 054.3 & 16.91 & 14.54 & 12.24 & 10.68 & 10.07 &  9.78 & 21.8 & XXV   & 24.1 pc\tablenotemark{a}\\
SCR 0138-5353      & 01 38 20.51 & -53 53 26.1 & 0.297 & 071.0 & 15.70 & 13.70 & 11.74 & 10.28 &  9.69 &  9.42 & 24.2 & XVIII & 39.7 pc\tablenotemark{a}\\      
SCR 0143-3840      & 01 43 03.26 & -38 40 07.5 & 0.188 & 101.2 & 12.71 & 10.51 &  9.22 &  8.52 &  7.85 &  7.68 & 19.3 & XXV   & \\
\enddata

\tablenotetext{a}{{2011AJ....141...21W}}
\tablenotetext{b}{{2006AJ....132.2360H}}
\tablenotetext{c}{Common proper motion; see Table~\ref{cpm-table}.}
\tablenotetext{d}{$\mu$ and/or PA suspect.}
\tablenotetext{e}{Not detected during automated search but noticed by eye during blinking process.}

\end{deluxetable}

\clearpage


\begin{deluxetable}{lcccccccccccll}
\rotate \tabletypesize{\scriptsize} \tablecaption{SCR Cool Subdwarf
  Candidates
\label{subdwarffive}}
\tablewidth{0pt}

\tablehead{\vspace{-8pt} \\
\colhead{Name}&
\colhead{RA}&
\colhead{DEC}&
\colhead{$\mu$}&
\colhead{$\theta$}&
\colhead{$B_J$}&
\colhead{$R_{59F}$}&
\colhead{$I_{IVN}$}&
\colhead{$J$}&
\colhead{$H$}&
\colhead{$K_s$}&
\colhead{Est Dist}&
\colhead{TSN}&
\colhead{Notes}\\

\colhead{}&
\colhead{(J2000)}&
\colhead{(J2000)}&
\colhead{($\arcsec$)}&
\colhead{($\degr$)}&
\colhead{}&
\colhead{}&
\colhead{}&
\colhead{}&
\colhead{}&
\colhead{}&
\colhead{(pc)}&
\colhead{}&
\colhead{}}

\startdata
\vspace{0pt} \\

SCR 0002-4644A & 00 02 35.66& -46 44 51.9 & 0.197 & 122.5 &  17.20 &  15.39 &  14.70 &  14.23 &  13.60 &  13.45 & [328.3]& XXV   &  \tablenotemark{a}  \\
SCR 0002-4644B & 00 02 36.20& -46 44 57.8 & 0.240 & 112.3 &  21.82 &  19.48 &  18.33 &  16.95 &  16.17 &  15.92 & [625.3]& XXV   &  \tablenotemark{a}  \\
SCR 0005-6152  & 00 05 31.98& -61 52 48.7 & 0.283 & 161.1 &  18.40 &  16.25 &  15.16 &  14.36 &  13.70 &  13.69 & [324.9]& XVIII &                     \\
SCR 0008-5843  & 00 08 15.37& -58 43 31.7 & 0.224 & 123.7 &  16.40 &  14.54 &  13.45 &  12.91 &  12.33 &  12.15 & [167.0]& XVIII &                     \\
SCR 0009-7305  & 00 09 48.15& -73 05 37.5 & 0.276 & 117.6 &  18.10 &  16.04 &  15.31 &  14.58 &  14.04 &  13.87 & [380.3]& XVIII &                     \\

\enddata

\tablenotetext{a}{Common proper motion system; see Table~\ref{cpm-table}.}
\tablenotetext{b}{Fewer than six relations for distance estimate, therefore unreliable [in brackets].}
\tablenotetext{c}{Coordinates not J2000.0 due to lack of proper motion or 2MASS data. SuperCOSMOS coordinates used instead.}
\tablenotetext{d}{Proper motion or position angle suspect.}

\end{deluxetable}

\clearpage


\begin{deluxetable}{lcccccccccccll}
\rotate \tabletypesize{\scriptsize} \tablecaption{SCR White Dwarf Candidates
\label{allscr-wd}}
\tablewidth{0pt}

\tablehead{\vspace{-8pt} \\
\colhead{Name}&
\colhead{RA}&
\colhead{DEC}&
\colhead{$\mu$}&
\colhead{$\theta$}&
\colhead{$B_J$}&
\colhead{$R_{59F}$}&
\colhead{$I_{IVN}$}&
\colhead{$J$}&
\colhead{$H$}&
\colhead{$K_s$}&
\colhead{Est. Dist. \tablenotemark{a}}&
\colhead{TSN}&
\colhead{Notes}\\

\colhead{}&
\colhead{(J2000)}&
\colhead{(J2000)}&
\colhead{($\arcsec$)}&
\colhead{($\degr$)}&
\colhead{}&
\colhead{}&
\colhead{}&
\colhead{}&
\colhead{}&
\colhead{}&
\colhead{(pc)}&
\colhead{}&
\colhead{}}

\startdata
\vspace{0pt} \\

SCR 0004-6120B& 00 04 45.41 & -61 23 40.0 & 0.171 & 127.6 & 16.86 & 16.76 & 16.53 & 16.43 & 15.93 & 16.47 & 59.3 & XVIII &      \\
SCR 0018-6851 & 00 18 08.56 & -68 51 19.4 & 0.220 & 091.6 & 16.55 & 16.46 & 16.48 & 16.62 & 16.13 & 17.18 & 52.2 & XVIII &      \\
SCR 0104-5742B& 01 04 12.14 & -57 42 48.6 & 0.239 & 091.1 & 16.19 & 15.89 & 15.78 & 15.67 & 15.56 & 15.75 & 34.5 & XVIII & DA6.5\tablenotemark{c}, 44.4 pc\tablenotemark{c}   \\
SCR 0125-4545 & 01 25 18.04 & -45 45 31.2 & 0.759 & 137.8 & 17.04 & 16.13 & 15.80 & 15.11 & 14.84 & 14.91 & 24.7 & XV    & DA8.5\tablenotemark{c},  24.9 pc\tablenotemark{c}   \\
SCR 0150-7207 & 01 50 38.49 & -72 07 16.8 & 0.334 & 223.9 & 16.18 & 15.71 & 15.24 & 15.65 & 15.64 & 15.42 & 28.0 & XVIII & DC\tablenotemark{c},  36.2 pc\tablenotemark{c}      \\
SCR 0245-6038 & 02 45 27.77 & -60 38 58.2 & 0.196 & 049.6 & 17.15 & 16.36 & 16.07 & 15.83 & 15.47 & 15.66 & 29.9 & XVIII &      \\
SCR 0252-7522 & 02 52 45.57 & -75 22 44.5 & 0.496 & 063.5 & 17.10 & 16.32 & 16.17 & 15.77 & 15.76 & 15.34 & 29.6 & XII   & DC\tablenotemark{b},  34.7 pc\tablenotemark{b}     \\
\enddata
\tablenotetext{a}{Estimate given using relation of \cite{2001Sci...292..698O}}
\tablenotetext{b}{\cite{2007AJ....134..252S}}
\tablenotetext{c}{\cite{2008AJ....136..899S}}
\tablenotetext{d}{No 2MASS data available}
\tablenotetext{e}{\cite{2009AJ....137.4547S}}
\end{deluxetable}

\clearpage


\begin{deluxetable}{lrrrlrrrrrl}
\rotate 
\tabletypesize{\scriptsize} 
\tablecaption{Common Proper Motion Systems 
\label{cpm-table}}
\tablewidth{0pt}
\tablehead{

\colhead{Primary}&
\colhead{$\mu$}&
\colhead{$\theta$}&
\colhead{Distance}&
\colhead{Companion(s)}&
\colhead{$\mu$}&
\colhead{$\theta$}&
\colhead{Distance}&
\colhead{Separation}&
\colhead{Position Angle}&
\colhead{notes}\\

\colhead{}&
\colhead{($\arcsec$)}&
\colhead{($\degr$)}&
\colhead{(pc)}&
\colhead{} &
\colhead{($\arcsec$)}&
\colhead{($\degr$)}&
\colhead{(pc)}&
\colhead{($\arcsec$)}&
\colhead{($\degr$)}&
\colhead{}}
\startdata

\vspace{-11pt} \\
\multicolumn{11}{c}{Probable Common Proper Motion Systems}\\
\tableline\\

BD-05 6069      & 0.228 &  123.8   &  36.3  & SCR 2352-0404B  & 0.213 &   128.5 &   237.0 &   531.0 &   1.1   & \tablenotemark{d}               \\
CF 11339        & 0.255 &  276.5   &  17.8  & SCR 1344-4535B  & 0.247 &   277.2 &    26.1 &   27.6  &   52.2  &                                 \\
G 077-057       & 0.241 &  173.3   &  91.4  & SCR 0329-0502B  & 0.226 &   180.5 &   264.9 &   15.3  &   78.4  & \tablenotemark{b}               \\
G 154-053       & 0.296 &  220.7   &  34.0  & SCR 1816-1246B  & 0.307 &   232.1 &    46.0 &   15.5  &   45.2  & \tablenotemark{b}               \\
HD 207554       & 0.189 &  100.8   &  24.0  & SCR 2150-1812B  & 0.182 &   113.3 &   111.6 &   57.4  &   10.5  & \tablenotemark{db}              \\
\enddata

\tablenotetext{a}{Primary detected by eye during blinking process.}
\tablenotetext{b}{Companion(s) detected by eye during blinking process.}
\tablenotetext{c}{Unresolved pair on plate; no proper motion, position angle, or distance estimate for at least one component.}
\tablenotetext{b}{Fewer than six relations for distance estimate, therefore unreliable [in brackets].}
\tablenotetext{e}{No $BRI$ photometry data available, therefore no distance estimate.}
\tablenotetext{f}{Proper motion or position angle suspect.}
\tablenotetext{g}{White dwarf candidate with unreliable distance [in brackets]. More accurate estimate in notes if available. See Table~\ref{allscr-wd}}
\tablenotetext{h}{Subdwarf candidate with unreliable distance [in brackets]. See Table~\ref{subdwarffive}}
\tablenotetext{i}{No 2MASS data available for component, therefore no distance estimate}
\tablenotetext{j}{Distance unreliable due to blended photometry}

\end{deluxetable}

\clearpage


\hoffset-00pt{}
\begin{deluxetable}{cccc}
\tablecaption{Distance Estimate Statistics for SCR Red Dwarf Systems\tablenotemark{a}
\label{diststats}}
\tablewidth{0pt}

\tablehead{\vspace{-8pt} \\
\colhead{Proper motion}&
\colhead{d $\leq$ 10 pc}&
\colhead{10 pc $<$ d $\leq$ 25 pc}&
\colhead{d $>$ 25 pc}}

\startdata

$\mu$ $\geq$ 1$\farcs$00 yr$^{-1}$                           & 2 $+$ 0 &  0 $+$ 0  &   6 $+$ 0 \\
1$\farcs$00 yr$^{-1}$ $>$ $\mu$ $\geq$ 0$\farcs$80 yr$^{-1}$ & 0 $+$ 0 &  3 $+$ 0  &   3 $+$ 0 \\
0$\farcs$80 yr$^{-1}$ $>$ $\mu$ $\geq$ 0$\farcs$60 yr$^{-1}$ & 1 $+$ 0 & 11 $+$ 0  &  48 $+$ 0 \\
0$\farcs$60 yr$^{-1}$ $>$ $\mu$ $\geq$ 0$\farcs$40 yr$^{-1}$ & 2 $+$ 0 & 24 $+$ 0  & 188 $+$ 0 \\
0$\farcs$40 yr$^{-1}$ $>$ $\mu$ $\geq$ 0$\farcs$18 yr$^{-1}$ & 2 $+$ 2 & 29 $+$ 77 & 1542 $+$ 2715  \\
\\
\tableline
\vspace{-3pt}
Total                                                       &  9 & 144 & 4502 \\

\enddata

\tablenotetext{a}{Entire SCR sample excluding white dwarf candidates
  and new common proper motion companions to known objects noticed by
  eye. The first number is the number of discoveries from previous
  papers. The second is the number of discoveries from the current
  paper. This paper did not search $\mu$ $>$ 0$\farcs$40 $yr^{-1}$ so
  no stars were added in those proper motion ranges.}

\end{deluxetable}

\clearpage



\figcaption[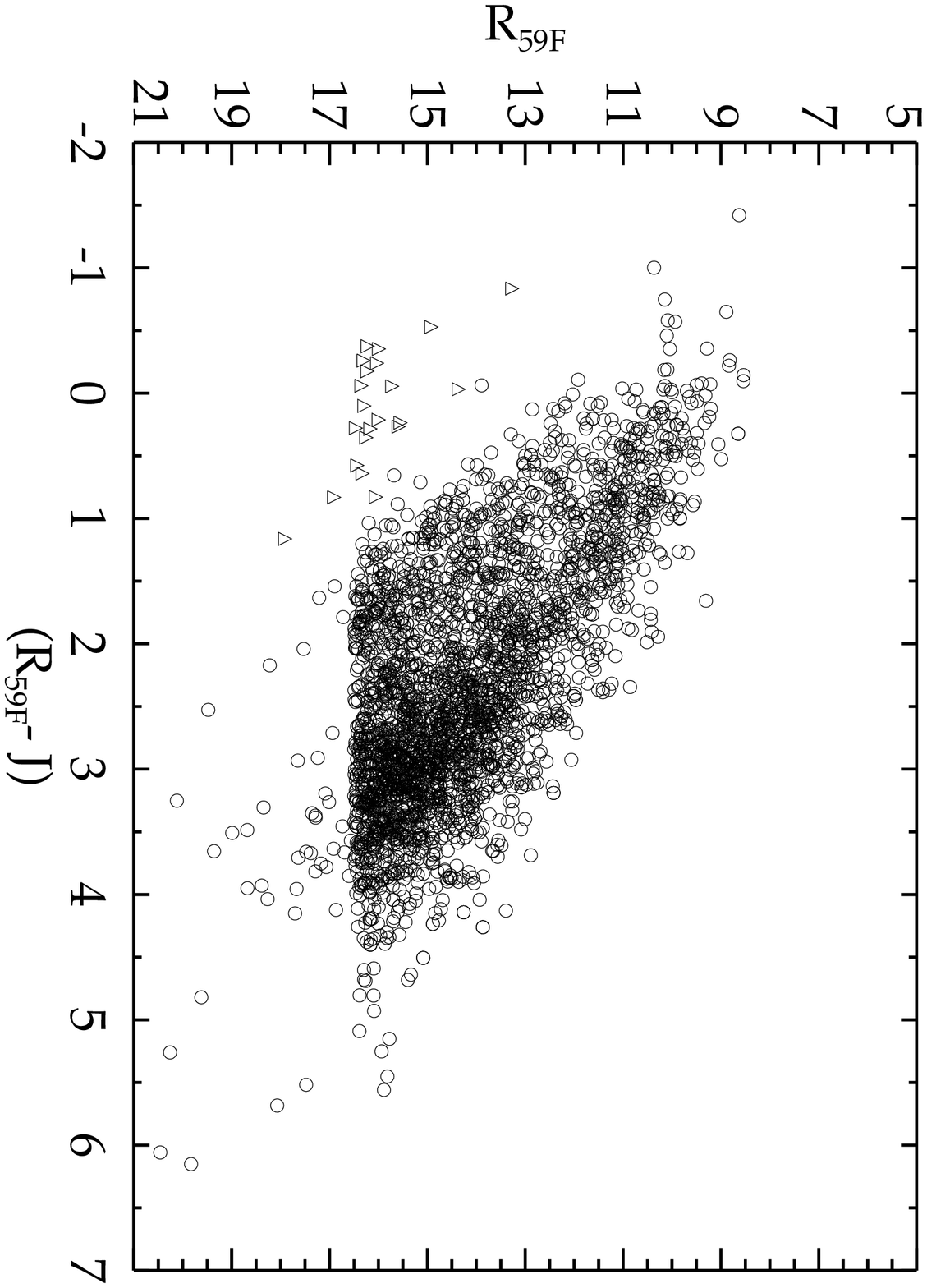]{Color-apparent magnitude diagram for new SCR
objects with 0$\farcs$40 yr$^{-1}$ $>$ $\mu$ $\ge$ 0$\farcs$18
yr$^{-1}$ found during the search described in this paper.  Data
points below $R_{59F}$ $=$ 16.5 are common proper motion companions
found during the blinking process. Triangles represent white dwarf
candidate objects from the RPM diagram.
\label{cmdnew}}


\figcaption[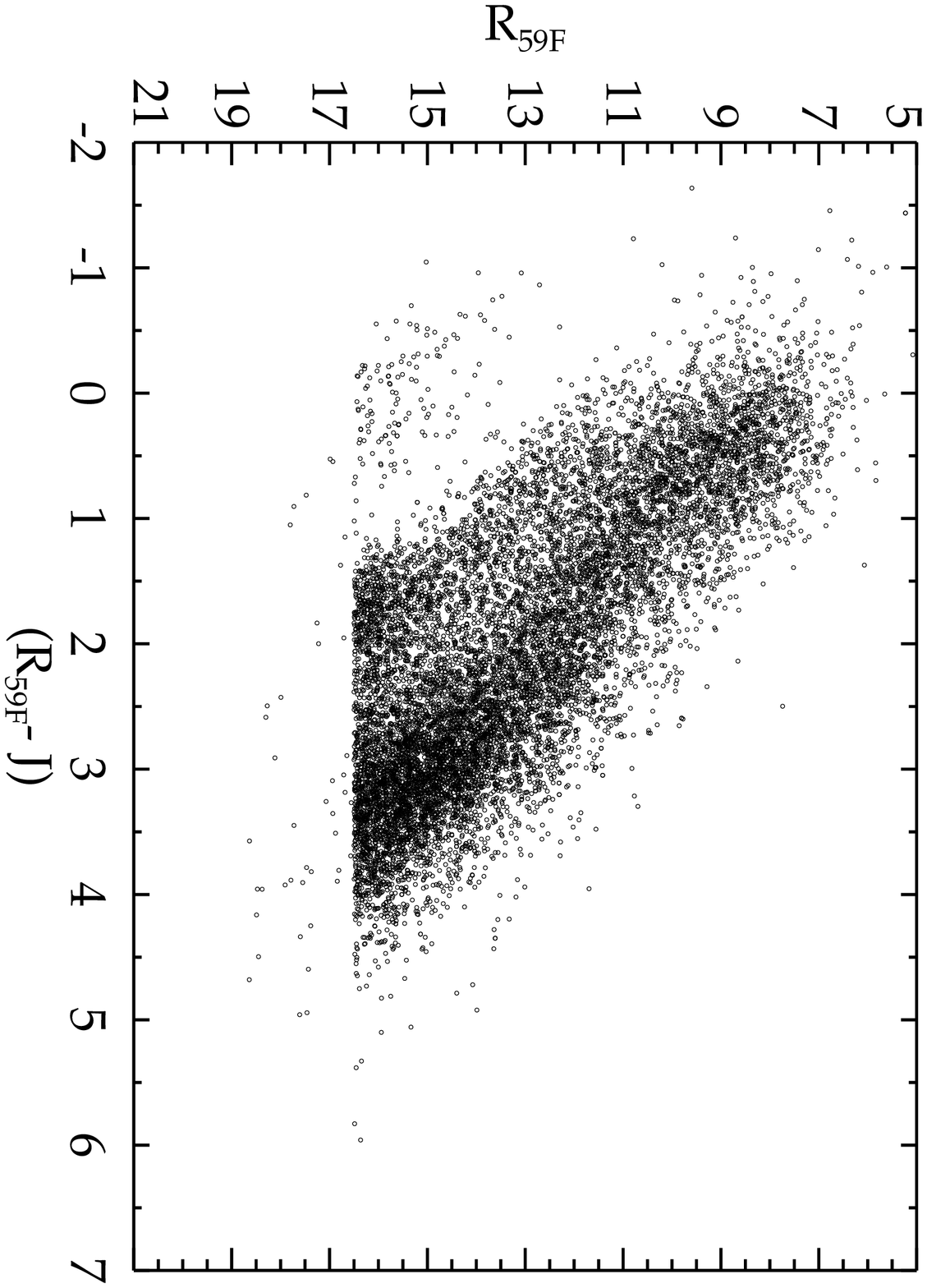]{Color-apparent magnitude diagram for known
objects with 0$\farcs$40 yr$^{-1}$ $>$ $\mu$ $\ge$ 0$\farcs$18
yr$^{-1}$ found during the search described in this paper.  Data
points below $R_{59F}$ $=$ 16.5 are common proper motion companions
found during the blinking process.
\label{cmdknown}}


\figcaption[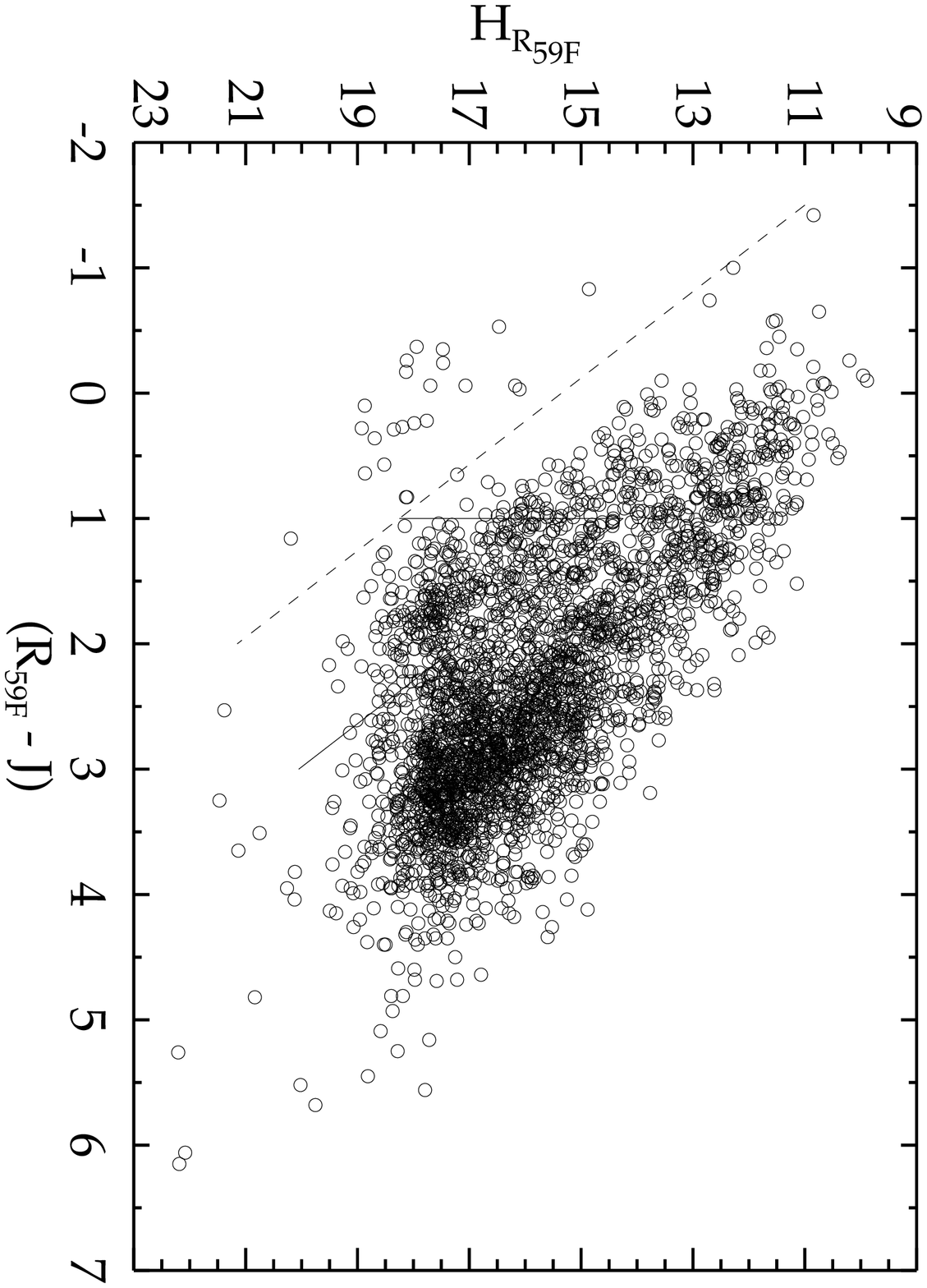]{Reduced proper motion diagram for new SCR
systems with 0$\farcs$40 yr$^{-1}$ $>$ $\mu$ $\ge$ 0$\farcs$18
yr$^{-1}$ found during the search described in this paper. The dashed
line separates candidate white dwarfs and subdwarfs and matches that
in other TSN papers. The solid lines denote the upper and bluest
boundaries of the cool subdwarf candidate section.
\label{rpmd}}


\figcaption[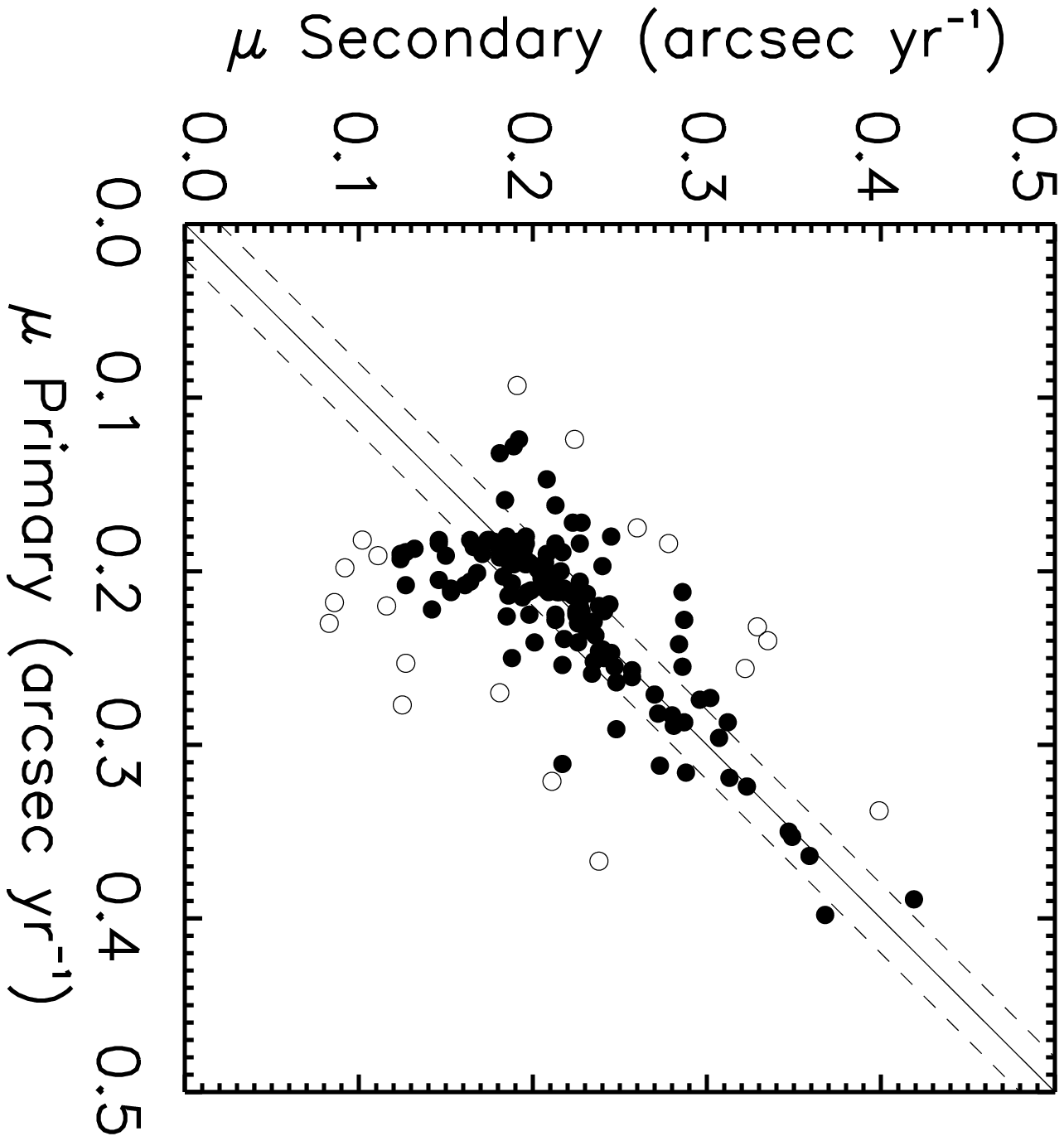]{Plot of the proper motion of the primary
versus that of its companion(s) in common proper motion systems. The
solid line denotes perfect agreement between the two proper motions,
while the dashed lines indicate limits of 0$\farcs$020 yr$^{-1}$ in
accordance with our uncertainties. Filled circles represent pairs in
which both members had data from the automatic phase of the
search. Open circles denote pairs in which proper motion data for at
least one component were gathered manually.
\label{mumu}}


\figcaption[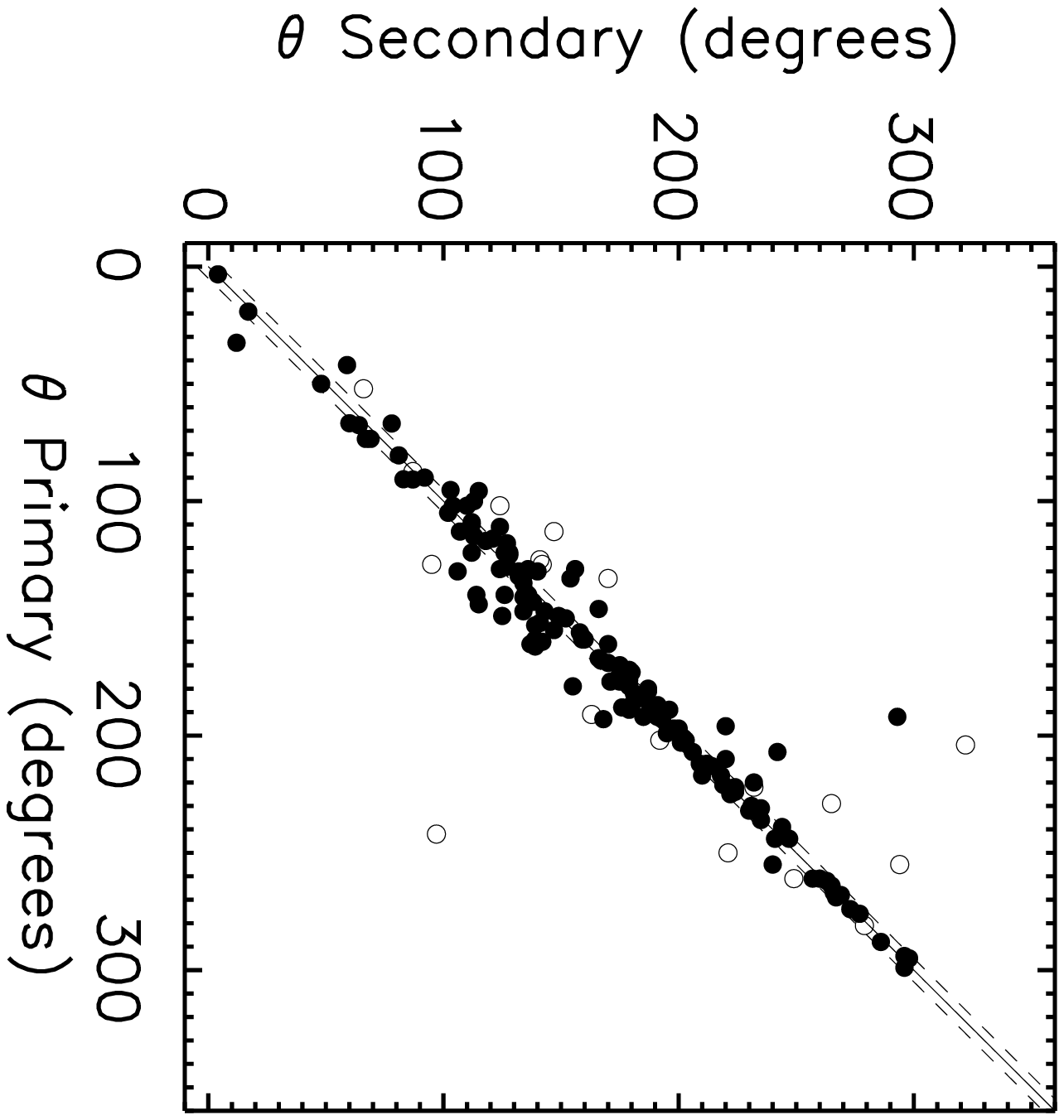]{Plot of the position angle of the primary's
  proper motion versus that of its companion(s) in common proper
  motion systems. The solid line denotes perfect agreement between the
  two, while the dashed lines indicate limits of 5$\degr$ in
  accordance with our uncertainties. Filled circles represent pairs in
  which both members had data from the automatic phase of the
  search. Open circles denote pairs in which position angle data for
  at least one component were gathered manually.
\label{papa}}


\figcaption[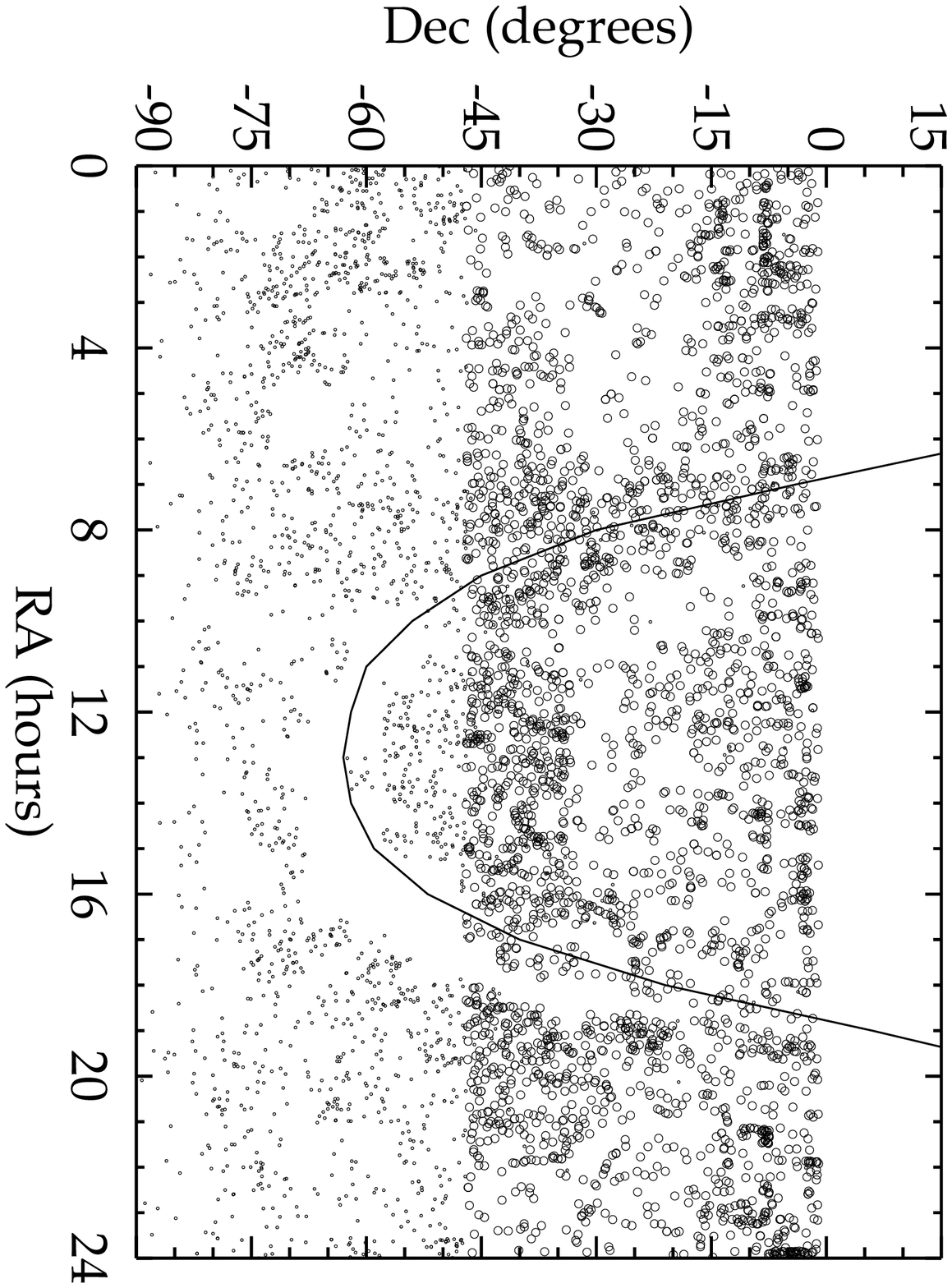]{Sky distribution of SCR systems. Large circles
represent discoveries from the current paper, while small circles
represent discoveries from previous SCR searches. The curve represents
the Galactic plane.
\label{distro}}


\figcaption[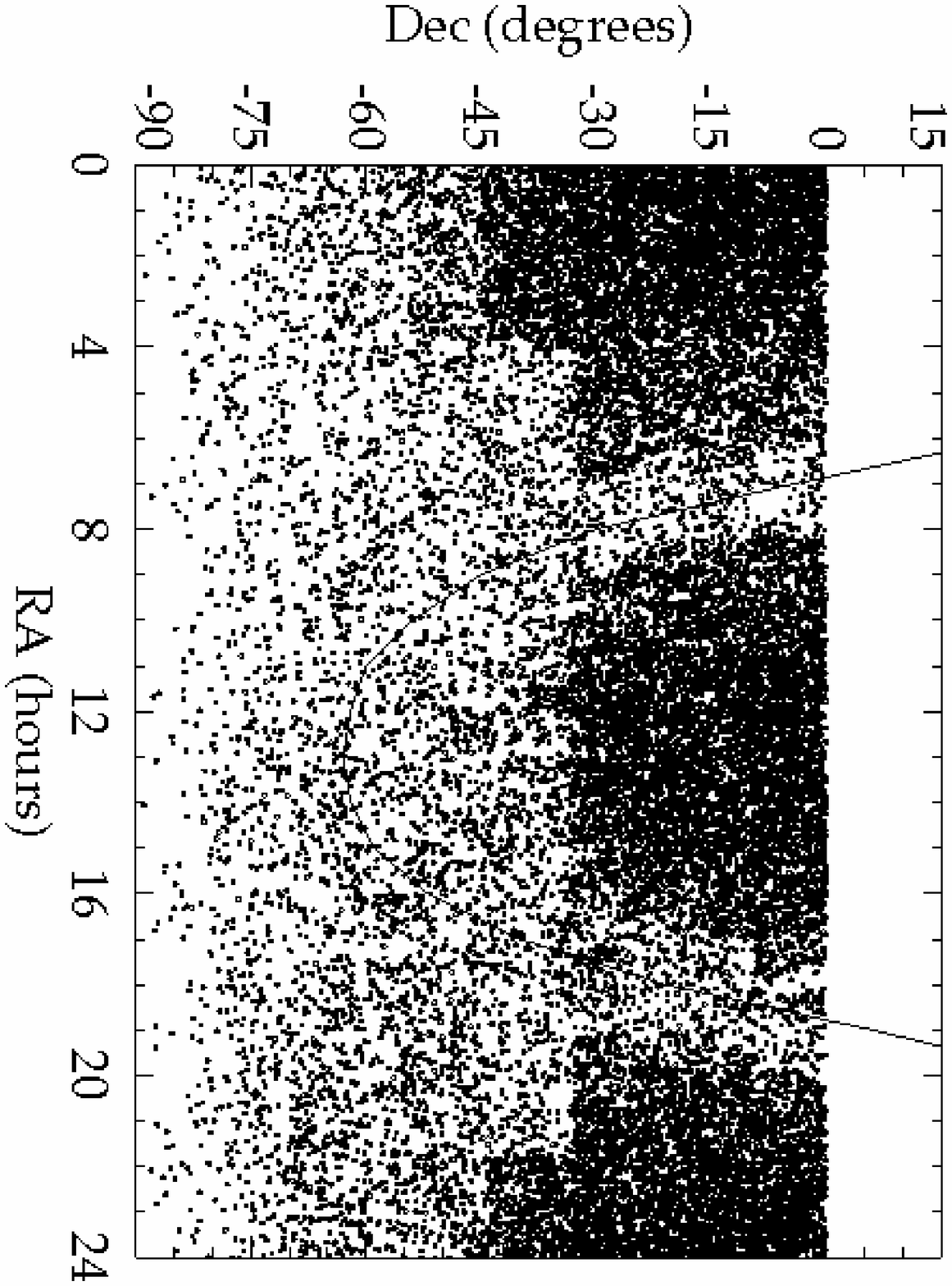]{Sky distribution of southern NLTT
systems. The curve represents the Galactic plane.}
\label{nlttdistro}

 
\begin{figure}
\plotone{fig1.eps}
\end{figure}

\begin{figure}
\plotone{fig2.eps}
\end{figure}

\begin{figure}
\plotone{fig3.eps}
\end{figure}

\begin{figure}
\plotone{fig4.eps}
\end{figure}

\begin{figure}
\plotone{fig5.eps}
\end{figure}

\begin{figure}
\plotone{fig6.eps}
\end{figure}

\begin{figure}
\plotone{fig7.eps}
\end{figure}

\clearpage


\end{document}